\documentclass[prl,twocolumn,showpacs,amsmath,amssymb]{revtex4-1}
\usepackage{graphicx}
\usepackage{dcolumn}
\usepackage{bm}
\begin{document}
\title{How Stress Can Reduce Dissipation in Glasses}
\author{Jiansheng Wu and Clare C. Yu}
\affiliation{Department of Physics and Astronomy, University of
California, Irvine CA 92697}
\date{\today}
\begin{abstract}

We propose that stress can decrease the internal friction of
amorphous solids, either by increasing the potential barriers of defects, thus reducing 
their tunneling and thermal activation that produce loss, or by decreasing the coupling 
between defects and phonons. This stress can be from impurities, atomic bonding 
contraints, or externally applied stress. Externally applied stress also reduces
mechanical loss through dissipation dilution. Our results are consistent with the 
experiments, and predict that stress could substantially reduce dielectric loss 
and increase the thermal conductivity.

\end{abstract}

\maketitle

\section{I. Introduction}

At low temperatures between 0.1 K and 10 K, a wide variety of
amorphous solids exhibit a universal plateau in their mechanical
dissipation $Q^{-1}\sim 10^{-4}-10^{-3}$ \cite{Zeller1971,Pohl2002}.
However, there are exceptions such as in amorphous silicon 
where doping with 1 at.\% of hydrogen reduces the low temperature
internal friction plateau by about a factor of 200 \cite{Liu1997}.
In addition the dissipation in high stress silicon nitride (Si$_3$N$_4$) 
thin films, which show no long range order in X-ray diffraction
and TEM images, is 2 to 3 orders of magnitude lower than in amorphous SiO$_2$ 
from 4 K up to room temperature \cite{Southworth2009}. Such a large effect 
is surprising since the stress of 1.2 GPa corresponds to only about 70 K/atom. 
Even the dissipation of stress relieved Si$_3$N$_4$ has a $Q$ that is about
an order of magnitude lower than typical amorphous solids \cite{Southworth2009}.

So far no theoretical explanation for these results has been presented.
In this paper we propose that all these reductions in dissipation are
due to stress but cannot be explained by one 
physical effect. Impurities, dopants, and internal bond constraints can produce
internal stress. Externally applied stress can reduce dissipation through 
dissipation dilution \cite{Saulson1990} as Saulson has pointed out 
\cite{PrivCommSaulson}. In addition we propose that stress, whether internal
or external, can reduce the dissipation
produced by microscopic defects known as two level systems (TLS), 
either by increasing TLS barrier heights or by decreasing the
coupling between phonons and TLS. Our goal is to urge experimentalists to
make further measurements to quantify the role of dissipation dilution as well
as to differentiate between these two possible effects of stress on TLS.  
 
In dissipation dilution \cite{Huang1998} externally
applied stress increases the stiffness of materials without increasing
their loss, resulting in a higher $Q$.
A simple example of dissipation dilution would be the increase in $Q$ of
a mass suspended from a lossy spring when a stiffer lossless spring is
added in parallel to the original spring. Since $Q=f_o/(\Delta f)$ where
$f_o$ is the resonant frequency and $\Delta f$ is the line width
(full width half max), $f_o$, and hence $Q$, increase without increasing 
the damping. In the appendix we estimate that 
a thin film square resonator of high stress silicon nitride could have a
$Q$ up to 40,000 times higher than a hypothetical stress-relieved silicon
nitride resonator due to dissipation dilution. This far exceeds the
experimental factor of order 150 by which external stress increases Q in
high stress silicon nitride \cite{Southworth2009}. The full enhancement of 
40,000 is not realized 
probably due to external sources of dissipation, e.g., clamping losses. 

Dissipation dilution only plays a role when there is externally applied stress.
So even though dissipation dilution can have a dramatic effect, it cannot explain
why dissipation is lowered by an order of magnitude or more
in materials which have no externally applied stress,
e.g., in silicon doped with 1 at.\% hydrogen \cite{Liu1997} or in stress
relieved Si$_3$N$_4$ \cite{Southworth2009}. Also, in addition to dissipation
dilution, external stress may reduce the internal friction arising from
microscopic defects. To understand this, we note that
$Q^{-1}=A\phi$ where $\phi$ is internal friction, and $A$ is due to dissipation
dilution and is a function of macroscopic parameters, e.g., elastic moduli
\cite{Huang1998}. (For the rest of the paper, except where noted otherwise, we
will focus on the internal friction and set $A=1$ so that we can use
$Q^{-1}$ in place of $\phi$ in order to be consistent with the accepted 
notation in the
field of glasses at low temperatures.) We propose
two possible ways in which stress could reduce
internal friction: either by increasing the barrier heights of microscopic 
fluctuating defects or by decreasing the coupling (deformation potential $\gamma$)
between phonons and two level systems (TLS). We fit existing, but incomplete,
experimental data on dissipation, specific heat, and thermal conductivity for
silicon nitride and SiO$_2$, finding somewhat better fits to the
$Q^{-1}$ data of Si$_3$N$_4$ at high temperatures with the barrier height model. 
Further measurements could distinguish between these two models.

In glasses at low temperatures, acoustic loss at low frequencies is 
attributed to TLS
\cite{Pohl2002, Phillips1972, Anderson1972, Jackle1972, Jackle1976, Hunklinger1986}.
While the microscopic nature of TLS is a mystery, one can think of a TLS 
as an atom or group of atoms in a double well potential that can sit 
in either well. At low temperatures, the lowest 2 energy 
levels dominate. The TLS density of states is assumed to be uniform at 
energies below a few Kelvin, so if stress merely shifts the density of states, 
there should be no effect. At low frequencies and temperatures, the primary 
mode of attenuation is relaxation in which the phonon at the measurement
frequency modulates the TLS energy level spacing 
\cite{Hunklinger1976}. The measurement frequency is not related to the
TLS energy because the
incident phonon can modulate TLS with any energy splitting. 
Attenuation occurs when 
the TLS population readjusts to the equilibrium Boltzmann distribution
with the aid of the entire thermal distribution of phonons. 

Low acoustic loss could have important implications for dielectric loss
since the two are completely analogous within the TLS model \cite{Hunklinger1981}.
TLS with electric dipole moments can produce
dielectric loss by attenuating photons. So we would expect stressed
dielectrics to also have low dielectric loss that could make them useful
substrates to reduce loss and noise in superconducting qubit circuits
\cite{Martinis2005}. For example, hydrogenating
amorphous silicon nitride decreases its dielectric loss tangent
by approximately a factor of 50 \cite{Paik2010}.

At low temperatures tunneling dominates, but
at higher temperatures thermal activation over energy barriers 
becomes important.
One possibility is that stress increases the potential energy barriers $V$
which reduces tunneling and thermal activation, thus effectively reducing
the number of defects and the internal friction. We will show that 
this approach is quantitatively consistent with measurements of $Q^{-1}$ in 
stress relieved Si$_3$N$_4$, and, 
even if we ignore dissipation dilution and demand that the entire reduction be
due to a reduction in internal friction, with measurements of $Q^{-1}$ in high 
stress Si$_3$N$_4$. We use a single set of parameters to calculate $Q^{-1}$, 
the specific heat $C(T)$, and the thermal conductivity $\kappa(T)$ 
in SiO$_2$ and silicon nitride. Since low dissipation implies a long phonon 
mean free path and a high thermal conductivity,
we predict that the thermal conductivity of stress relieved
Si$_3$N$_4$ is an order of magnitude higher than amorphous SiO$_2$ from 4 K up to
room temperature, and, if there is no dissipation dilution, 
the thermal conductivity of high stress Si$_3$N$_4$
could be even higher, potentially making silicon
nitride a useful substrate for integrated circuits where cooling is important. 

The paper is organized as follows. We describe our calculations of the
dissipation, thermal conductivity, specific heat, and dielectric loss
in section II. In section III, we explain our procedure for determining
the parameters for fitting the experimental data. The results of those
fits to the specific heat, thermal conductivity, and dissipation are
presented in section IV. We discuss why the dissipation of stress
relieved Si$_3$N$_4$ is lower than ordinary materials in section V.
We discuss the possibility that stress could reduce dielectric loss in 
section VI. In Section VII we present an alternative model for how
stress could lower the dissipation, namely by reducing the coupling
between TLS and phonons. We summarize our work in section VIII.

\section{II. Calculations of Dissipation, Thermal Conductivity, Specific Heat, and
Dielectric Loss}

Let us briefly review the TLS model \cite{Anderson1972,Phillips1972}.
The TLS Hamiltonian is $H=H_{o}+H_{e}$ where
$H_{o}=(1/2)\left[\Delta\sigma_{z}-\Delta_{o}\sigma_{x}\right]$ and
$H_{e}=\gamma  e \sigma_z$ where $\Delta$ is the energy asymmetry
between the potential energy wells, $\Delta_o$ is the tunneling matrix element,
$\gamma$ is the deformation potential,
$e$ is the strain field, and $\sigma_x$ and $\sigma_z$ are Pauli matrices. 
The energy eigenvalues of $H_o$ are $E=\pm \sqrt{\Delta_0^2+\Delta^2}$.
We follow Tielburger {\it et al.} \cite{Tielburger1992} and 
approximate the double well by two overlapping
harmonic oscillator wells, each with energy level spacing $\hbar\Omega_{o}$. 
The tunneling matrix element
$\Delta_{o}$ is given by the WKB approximation \cite{Tielburger1992}:
\begin{equation}
    \Delta_0 =\frac{\hbar\Omega_{o}}{\pi}\left(\sqrt{\Lambda+1}+\sqrt{\Lambda}\right)
              \exp(-\sqrt{\Lambda^2+\Lambda})
\label{eq:WKB}
\end{equation}
where $\Lambda=2V/(\hbar\Omega_{o})$ and $V$ is the height of the energy barrier.
Fits to the low temperature thermal conductivity find
the TLS density of states $\bar P$ that couples to phonons to be 
approximately constant. However, an excess of local vibrational states, 
referred to as the
boson peak, is evident at higher temperatures and energies
\cite{Zeller1971,BosonPeak}. We 
model these modes by Einstein oscillators with a step function
in the density of states that starts at an energy $E_o$ typically between
10 and 40 K \cite{Yu1987}. 

According to the TLS model, at low frequencies ($\nu<$ 1 THz) and 
low temperatures (0.1 K $< T <$ 10 K), $Q^{-1}$ is a temperature 
independent constant given 
by \cite{Jackle1972}:
\begin{equation}
Q^{-1}_{o}=\frac{\pi {\bar P}\gamma^2}{2 \rho v^2},
\label{eq:Qo}
\end{equation}
where $\rho$ is the mass density, and $v$ is the sound velocity. 
The sources of attenuation are TLS relaxation processes 
($Q^{-1}_{\rm rel, TLS}$), resonant scattering of phonons 
from TLS ($Q^{-1}_{\rm res, TLS}$) and Einstein oscillators 
($Q^{-1}_{\rm EO}$) in which the phonon energy matches the energy level 
spacing, and Rayleigh scattering ($Q^{-1}_{\rm Ray}$)
from small scatterers of size $a$ such that $ka\stackrel{<}{\sim}1$ where
$k$ is the phonon wavevector \cite{Yu1987}. Yu and Freeman
\cite{Yu1987} found that $a=k^{-1}=\hbar v/E_o$ is consistently $\sim$ 25\% 
larger than the size \cite{Freeman1986} of a molecular unit for 
SiO$_2$, GeO$_2$, polystyrene, and PMMA (polymethylmetahcrylate). 
Just as in their work, we cut off Rayleigh scattering at $E_o$. 
We include thermal activation as well as direct phonon
relaxation in the TLS relaxation processes \cite{Tielburger1992}, 
and assume that the relaxation attenuation from Einstein oscillators is negligible
\cite{Yu1987}. 
Thus we can write \cite{Yu1987,Tielburger1992}
\begin{eqnarray}
 Q^{-1} &=& \left\{\begin{array}{ll} Q^{-1}_{\rm res, TLS}+Q^{-1}_{\rm rel, TLS}+
Q^{-1}_{\rm Ray}, & E<E_{o} \\
Q^{-1}_{\rm res, TLS}+Q^{-1}_{\rm rel, TLS}+Q^{-1}_{\rm EO}, & E>E_{o}.
  \end{array} \right.
\label{eq:Qinv}
\end{eqnarray}
The attenuation due to TLS relaxation is given by
\begin{equation}
Q^{-1}_{\rm rel, TLS} = \frac{2 Q_0^{-1}}{\pi k_{\rm B} T} \int_{V,\Delta}
\left(\frac{\Delta}{E}\right)^2 {\rm sech}^2\frac{E}{2 k_B T}
\frac{\omega\tau}{1+(\omega\tau)^2}
\label{eq:Qrel}
\end{equation}
where
$\int_{V,\Delta} \equiv  \int_{0}^{V_{\rm max}} dV \int_{0}^{2V}
d\Delta P(\Delta,V)/{\bar P}$ with $V_{\rm max}=V_0+6\sigma_0$.
$P(\Delta,V)$ is the TLS distribution of $\Delta$ and 
$V$. We assume that $\Delta$ has a uniform distribution
and $V$ has a Gaussian distribution with an average $V_0$ and a variance
$\sigma^{2}_0$ \cite{Tielburger1992}:
\begin{equation}
    P(\Delta,V)=\frac{2\bar P}{\hbar\Omega_o}
                \exp\left[-\frac{(V-V_0)^2}{2\sigma_0^2}\right].
\label{eq:PV}
\end{equation}

The TLS relaxation rate $\tau^{-1}$ is the sum of the direct phonon relaxation 
rate $\tau_{\rm d}^{-1}$ in which the excited TLS decays to the
ground state by emitting a phonon, and the rate $\tau^{-1}_{\rm Arr}$
of Arrhenius activation over the barrier:
\begin{eqnarray}
 \tau^{-1}&=&\tau^{-1}_{\rm d}+\tau^{-1}_{\rm Arr} \label{eq:tau}\\
 \tau^{-1}_{\rm d} &=& \sum_{a=\ell,t} \left(\frac{\gamma_a^2}{v^5_a}\right)
\frac{E\Delta_0^2}{2\pi\rho\hbar^{4}}\coth\left(\frac{E}{2k_B T}\right) 
\label{eq:tau_d}\\
\tau^{-1}_{\rm Arr}&=& \tau_0^{-1}\cosh\left(\frac{\Delta}{2k_B T}\right)
e^{-V/k_B T}
\label{eq:tau_Arr}
\end{eqnarray}
where the sum is over the longitudinal and transverse phonon modes and
$\tau_0=2/\Omega_0$. For SiO$_2$ $\tau_0=4\times 10^{-12}s$. 
For $\omega\tau_m\ll 1$, $Q^{-1}_{\rm rel, TLS}\approx Q_0^{-1}$, where
$\tau_m$ is the minimum relaxation time for a TLS with energy $E$ at 
temperature $T$ \cite{Yu1987}. The Rayleigh and resonant phonon scattering 
terms are given by
\begin{eqnarray}
Q^{-1}_{\rm Ray} &=& B v {\omega}^3 \\
Q^{-1}_{\rm EO}&=& Q_0^{-1}\frac{2 S_\kappa}{\pi}\label{Q_EO}\\
Q^{-1}_{\rm res, TLS} &=& 2Q_0^{-1}\int_{V,\Delta}\tanh\frac{\hbar\omega}{k_{B}T}
\left(\frac{\Delta_0}{E}\right)^2\delta(E-\hbar\omega)\label{eq:Q_res}
\end{eqnarray}
where $S_\kappa$ is the step height in the density of states of the Einstein 
oscillators that is used to fit the thermal conductivity $\kappa$,
and $B$ is a constant.

$Q^{-1}$ is measured at low frequencies of order 1 MHz. 
Estimating the order of magnitude of the various contributions at 1 MHz and
1 K using the values of the parameters in Table 1 for SiO$_2$ (transverse
phonon modes), we find 
$Q^{-1}_{\rm rel, TLS}\sim Q^{-1}_{o}\sim 6\times 10^{-4}$, 
$Q^{-1}_{\rm res, TLS}\sim Q^{-1}_{o}\tanh(\hbar\omega/2k_BT)\sim 1 \times 10^{-8}$, 
and $Q^{-1}_{\rm Ray}\sim 2\times 10^{-15}$. 
Thus TLS relaxation dominates $Q^{-1}$ at low temperatures and low frequencies where
the plateau in $Q^{-1}$ is given by
\begin{equation}
Q^{-1}_{\rm plat}=Q^{-1}_o \exp\left[-\frac{V_0^2}{2\sigma_0^2}\right]
\label{eq:QlowT}
\end{equation}
This replaces Eq.~(\ref{eq:Qo}), and is obtained by plugging Eq.~(\ref{eq:Qrel}) into
Eq.~(\ref{eq:Qinv}) and noting that the dominant contribution to the
integral in Eq.~(\ref{eq:Qinv}) is for $V\ll V_0$ due to the exponential 
dependence of $\tau_{Arr}^{-1}$ and $\tau_{d}^{-1}$ on $V$. The factor of 
$\exp[-(V_0/\sigma_0)^2]$ in $P(\Delta,V)$ effectively reduces the number of 
active TLS. 

The relaxation time $\tau$ in Eq.~(\ref{eq:Qrel}) for $Q^{-1}_{\rm rel, TLS}$
is exponentially sensitive to the barrier height $V$ because both the tunneling matrix
element $\Delta_o$ in $\tau_{\rm d}$ (see Eqs.~(\ref{eq:WKB}) and (\ref{eq:tau_d})) 
and the thermal activation time
$\tau_{\rm Arr}$ given by Eq.~(\ref{eq:tau_Arr}) depend exponentially on $V$.
We assume that stress increases the barrier heights $V$, thus increasing 
the relaxation times $\tau_{\rm d}$ and $\tau_{\rm Arr}$, and reducing the
dissipation $Q^{-1}\approx Q^{-1}_{\rm rel, TLS}$. In our model stress
increases the average barrier height $V_0$ and decreases the variance
$\sigma_0^2$ in $P(\Delta,V)$.

In order to determine the values of the parameters required to fit $Q^{-1}$,
we need to fit the thermal conductivity $\kappa(T)$ and the specific heat $C(T)$.
The equations for $C(T)$ and $\kappa(T)$ are as follows.
In glasses heat is carried by phonons \cite{Zaitlin1975}.
$\kappa(T)$ is given by 
\begin{equation}
\kappa(T)=\frac{1}{3}\int_0^{\omega_D} C_{\rm D}(T,\omega)v \ell(T,\omega)d\omega
\label{eq:kappa}
\end{equation}
where $\omega_{\rm D}$ is the Debye frequency, and we approximate the 
phonon specific heat by the Debye specific heat $C_{\rm D}(T,\omega)$. The phonon
mean free path $\ell$ is related to $Q$ by 
\begin{equation}
\ell(T,\omega)=Q(T,\omega)v/\omega=Q(T,\omega)\lambda/(2\pi) 
\end{equation}
where $\lambda$ is the phonon wavelength.

The specific heat $C(T)$ has contributions from the phonons which we approximate
with the Debye specific heat $C_{\rm D}$, from TLS $C_{\rm TLS}$, and from local modes
which we model with Einstein oscillators $C_{\rm EO}$ \cite{Yu1987}:
\begin{equation}
C(T)=C_{\rm D}(T) + C_{\rm TLS}(T) + C_{\rm EO}(T) 
\end{equation}
where
\begin{eqnarray}
C_{\rm D} &=& 9 \frac{N}{V} k_B \left(\frac{T}{\Theta_D}\right)^3 
\int_0^{x_D} d x 4 x^4 \frac{e^x}{(e^x - 1)^2} \\
C_{\rm TLS} &=& k_B\bar{P}\int_{V,\Delta} x^2 \frac{e^x}{(e^x+1)^2} =\frac{\pi^2}{6}n_0 k_B^2 T \\   
C_{\rm EO}
&=&  n_o S_c k_B^2 T\int_{x_0}^{x_D} dx \frac{x^2 e^x}{(e^x-1)^2}
\end{eqnarray}
where $x=E/k_{\rm B}T$, $x_0=\hbar\Omega_0/k_BT$, $x_D=\Theta_D/T$, 
$N/V$ is the number density of formula units, and
$\theta(E)$ is a step function. $\Theta_D$ is the Debye temperature. 
$n_o$ is the TLS density of states that contributes to the specific heat,
and $S_{C}$ is the size of the step in the density
of states due to the Einstein oscillators that contribute to $C(T)$.

The dielectric loss tangent $\tan\delta$
is analogous to the acoustic dissipation $Q^{-1}$. At high frequencies and low 
temperatures ($\stackrel{<}{\sim}$ 1 K), the dominant scattering is resonant 
scattering of photons by TLS in which the photon energy matches the TLS
energy splitting. If the electromagnetic intensity $J$ is much less than 
the critical intensity $J_c$, we are below saturation, and can use 
Eq.~(\ref{eq:Q_res}) with $\tan\delta$
replacing $Q^{-1}_{res,TLS}$, and $Q^{-1}_o$ replaced by \cite{Schickfus1977}
\begin{equation}
Q^{-1}_{o,dielectric}=\frac{4\pi^{2}n_e p^2}{3\varepsilon_o\varepsilon_r}
\label{eq:dielectricTan}
\end{equation}
where $n_e$ is the density of TLS with electric dipole moments, $p$ is the
electric dipole moment, $\omega$ is the angular frequency of the incident photons,
$\varepsilon_o$ is the permittivity of the vacuum, and $\varepsilon_r$ is the
dielectric constant. Since the integral in Eq. (\ref{eq:Q_res}) is 
dominated by $V\ll V_0$, we can make the approximation 
\begin{equation}
\tan\delta=Q^{-1}_{o,dielectric}\exp\left[-\frac{V_0^2}{2\sigma_o^2}\right]
\tanh\left(\frac{\hbar\omega}{2k_BT}\right)
\end{equation}
for the barrier height model. For the model, described in Section VII,
where the stress modifies the
deformation potential $\gamma$, $V_0=0$, and
\begin{equation}
\tan\delta=Q^{-1}_{o,dielectric}\tanh\left(\frac{\hbar\omega}{2k_BT}\right)
\end{equation}

\section{III. Procedure for Fitting the Experimental Data}
\subsection{a. SiO$_2$}
To fit the data for SiO$_2$ we follow Tielburger {\it et al.} 
\cite{Tielburger1992} and set $V_0=0$. Then by fitting the low temperature 
plateau of $Q^{-1}$ using Eqs.~(2) and (12), we obtain
${\bar P}\gamma^2$. Fitting $Q^{-1}$ over the whole range of temperature
yields $\sigma_0$. The temperature of the rise in $Q^{-1}$ determines 
$\hbar\Omega_o/2$. Since $V_0$ and $\sigma_0$ are known, we can determine $n_0$ by 
fitting the specific heat $C(T)$ which then gives the value of ${\bar P}$. 
$\bar P$, $\gamma$, $V_0$, and $\sigma_0$
determine the low temperature thermal conductivity $\kappa(T)$ without any 
adjustable parameters. We set the energy $E_o$ of the
onset of the step in the density of states by 
$E_o=\Theta_{D}/(2\pi \times 1.27)$ \cite{Yu1987} where
$\Theta_{D}$ is the Debye temperature. 
Fitting $C(T)$ at higher temperatures determines the step $S_C$ in the density of
states due to local modes (Einstein oscillators). 
The fit to $\kappa(T)$ at high temperatures
gives the step in the density of states $S_{\kappa}$ and
the Rayleigh scattering parameter $B$.
Note that ${\bar P} < n_o$ and $S_{\kappa} < S_C$ because
not all of the degrees of freedom that contribute to the
specific heat scatter the phonons that are responsible for the
thermal conductivity. No one has tried before to see if one set of parameters 
can be used to fit the data for all these quantities.

\subsection{b. Si$_3$N$_4$}
Fitting the data for Si$_3$N$_4$ is complicated by the fact that
measurements of $Q^{-1}(T)$, $\kappa(T)$,
and $C(T)$ have not been made for the same stoichiometry of silicon nitride.
We assume a $\bar{P}\gamma^2$ value (such that $Q_0^{-1}\approx 10^{-4}-10^{-3}$)
for Si$_3$N$_4$ 
and use Eq.~(12) to fit the $Q^{-1}(T)$ data for 
low stress Si$_3$N$_4$ \cite{Southworth2009} to obtain $V_0$ and $\sigma_0$.
Assuming SiN$_{1.15}$ has the same values of $V_0$ and $\sigma_0$ as low 
stress Si$_3$N$_4$, we can obtain $n_0$ and $\bar P$ by fitting the $C(T)$ 
data of SiN$_{1.15}$. By fitting the $\kappa(T)$ data of SiN$_{1.15}$,
we obtain $\bar{P}\gamma^2$, and thus $\gamma$. Assuming Si$_3$N$_4$ has
the same  $\gamma$ as SiN$_{1.15}$, we obtain $\bar P$ for Si$_3$N$_4$. 
If this value is reasonable compared to the $\gamma$ values for SiN$_{1.15}$ 
and a-SiO$_2$, we stop. 
Otherwise, we choose another $\bar{P}\gamma^2$ and repeat the above procedure 
until we obtain a reasonable value of $\bar{P}$. 
We then follow the procedure given in the previous paragraph 
to fit $C(T)$ and $\kappa(T)$ for SiN$_{1.15}$ at higher temperatures 
to obtain the values of $B$, $S_{\kappa}$, and $S_{C}$. 

\section{IV. Results: Fits to Experimental Data}
\subsection{a. SiO$_2$}
\begin{figure}
\includegraphics[width=3.3in]{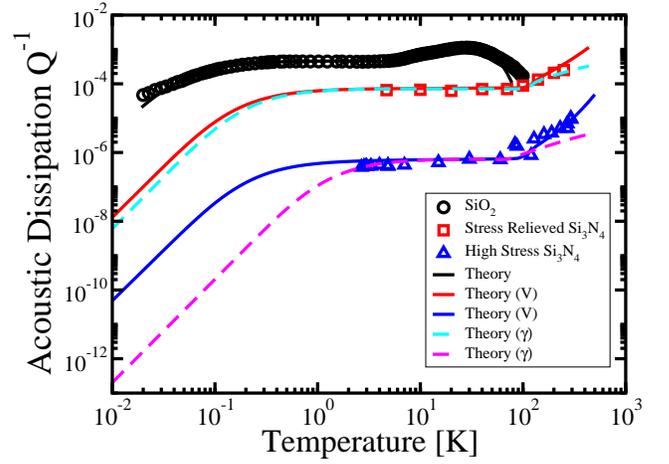}
\caption{(Color online)
Dissipation $Q^{-1}$ vs. $T$ for stress relieved Si$_3$ N$_4$ measured
at 3.5387 MHz (solid squares)\cite{Southworth2009}, high
stress Si$_3$N$_4$ measured at 1.526445 MHz (solid triangles)
\cite{Southworth2009}, and amorphous SiO$_2$ (open circles) measured at 
11.4 kHz \cite{Tielburger1992}. Solid lines are theoretical fits using the
model where stress reduces barrier height.
Dashed lines are our theoretical predictions associated with reducing $\gamma$.
Dissipation dilution factor $A=1$ in the theoretical curves.
}
\label{fig:Q}
\end{figure}

Our fits to the data for $\kappa(T)$, $C(T)$, and $Q^{-1}$
for SiO$_2$ and silicon nitride are shown in Figures \ref{fig:SpecHeat},
\ref{fig:thermCond} and \ref{fig:Q}
with the parameters given in Table 1. 
No one has tried before to see if one set of parameters
can be used to fit the data for all these quantities.
The fits to the SiO$_2$ data show that this can be done.

\begin{figure}[!h] \includegraphics[width=3.3in]{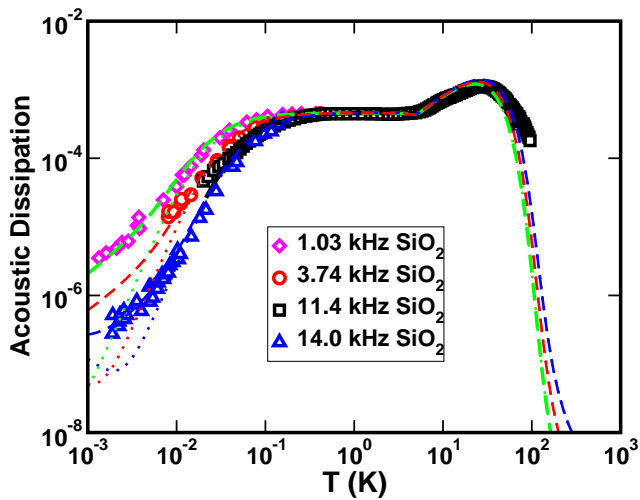}
  \caption{(Color online)
Acoustic dissipation $Q^{-1}$ vs. temperature for SiO$_2$ at various
frequencies. The data for SiO$_2$ at 11.4 kHz is from \cite{Tielburger1992}
while the rest of the SiO$_2$ data is from \cite{Fefferman2008}.
The fits to the
SiO$_2$ data with the linear term $bT$ are given by dashed lines while
the fits without the linear term added are shown as dotted lines.
}
\label{fig:invQbT}
\end{figure}

Fefferman {\it et al.} \cite{Fefferman2008} have reported that around 10 mK,
the acoustic dissipation of SiO$_2$ is linear in temperature. This linear
temperature dependence is attributed to interactions between TLS
\cite{Fefferman2008}.
In Figure \ref{fig:invQbT} we show our fits to the data from
\cite{Fefferman2008}. To obtain these fits we followed Fefferman
{\it et al.} \cite{Fefferman2008} and added a linear term
$\tau_{int}=bT(\Delta_0/E)^2$ \cite{Yu2004} where $b$ is a constant to
the expression for the relaxation rates in Eq.~(\ref{eq:tau}).
$b$ is a constant.

\subsection{b. Si$_3$N$_4$}
\begin{figure}
  \includegraphics[width=3.1in,angle=270]{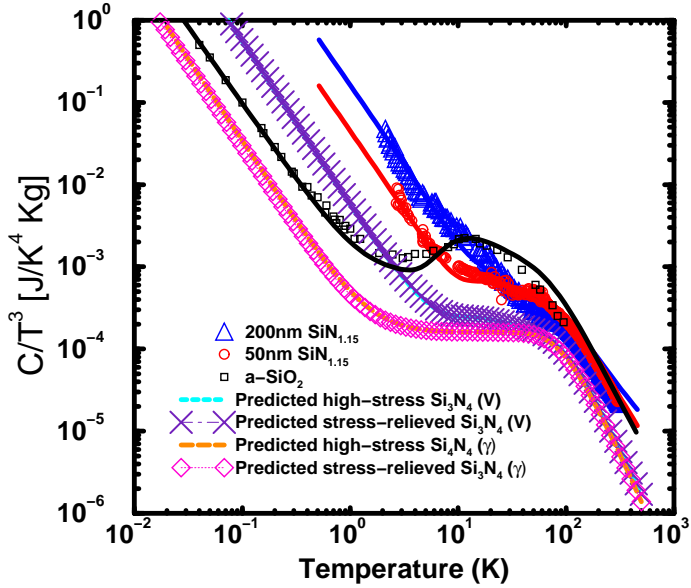}
  \caption{(Color online)
$C(T)/T^3$ vs.~$T$ for amorphous SiO$_2$ and silicon
nitride. Experimental data points are shown for 50 nm and 200 nm thick
SiN$_{1.15}$ \cite{Queen2009} and SiO$_2$. The SiO$_2$ $C(T)$ data
are from \cite{Zeller1971,Lasjaunias1975}. The solid lines through the points
are theoretical fits. Our predictions where stress affects $V$
or $\gamma$ are indicated in the legend by (V) and ($\gamma$), respectively.
$C(T)/T^3$ curves for high stress and stress relieved Si$_3$N$_4$ lie on
top of each other for the barrier height model. Similarly for the $\gamma$ model.
}
\label{fig:SpecHeat}
\end{figure}

\begin{figure}  
\includegraphics[width=3.1in,angle=270]{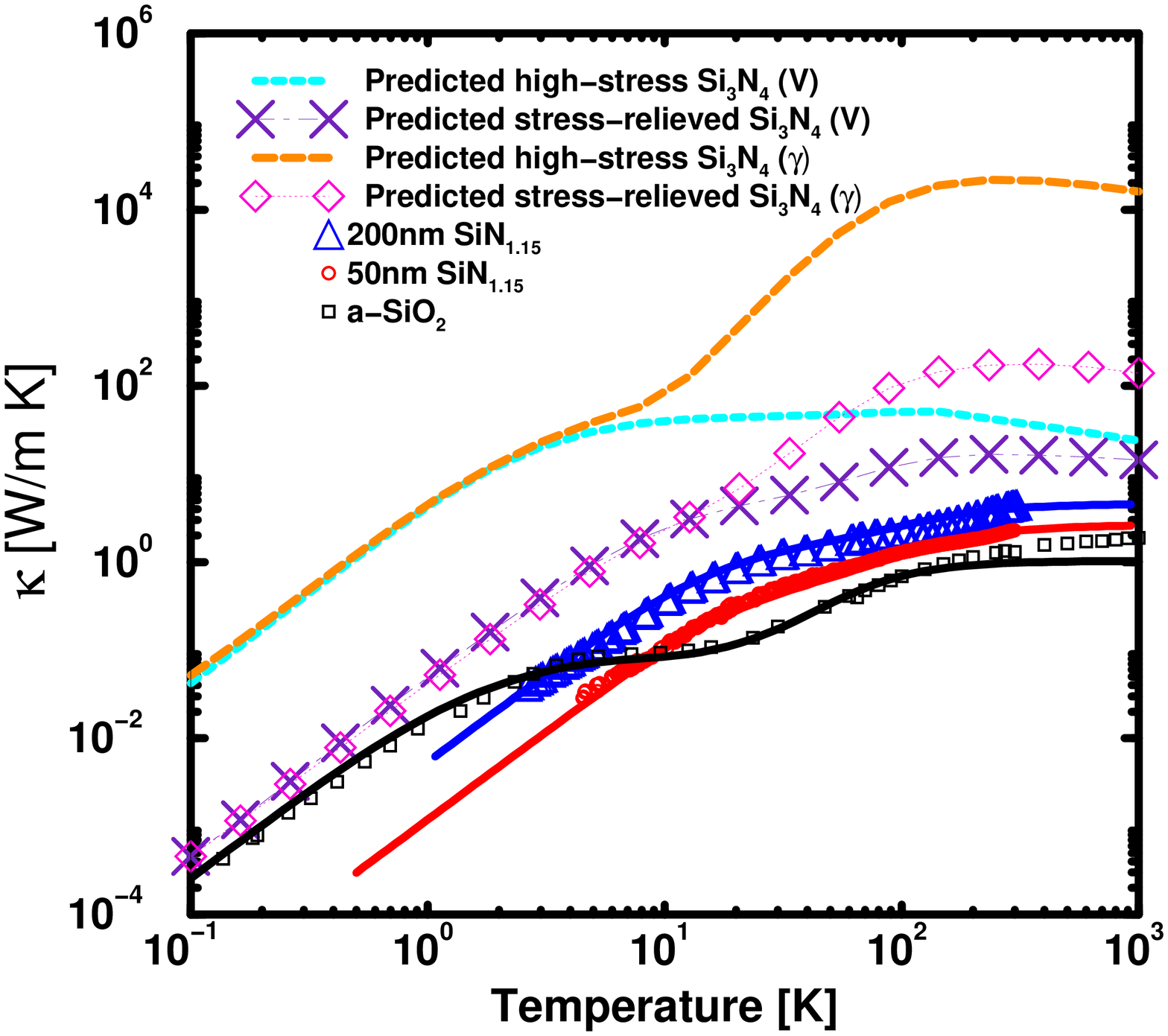}  
\caption{(Color online)
$\kappa(T)$ vs.~$T$ for amorphous SiO$_2$ and silicon
nitride. Experimental data points are shown for 50 nm and 200 nm thick
SiN$_{1.15}$ \cite{Queen2009} and SiO$_2$. The SiO$_2$ $\kappa(T)$
data are from \cite{Smith1975,Touloukian1970}. The solid lines through the points
are theoretical fits. Our predictions where stress affects $V$
or $\gamma$ are indicated in the legend by (V) and ($\gamma$), respectively.
At low temperatures $\kappa(T)$ for high stress Si$_3$N$_4$ is the same for
the $V$ and $\gamma$ models. Similarly for stress relieved Si$_3$N$_4$.
}
\label{fig:thermCond}
\end{figure}

Fitting the data for Si$_3$N$_4$ is complicated by the fact that
measurements of $Q^{-1}(T)$, $\kappa(T)$,
and $C(T)$ have not been made for the same stoichiometry of silicon nitride.
Assuming that $Q^{-1}=\phi$, i.e., with no dissipation dilution, our predictions 
for $C(T)$ and $\kappa(T)$ for high stress and stress-relieved
Si$_3$N$_4$ are shown in Figs.~\ref{fig:SpecHeat} and \ref{fig:thermCond}. 
Around 3 K, $\kappa(T)$ for 
stress relieved Si$_3$N$_4$ is about an order of magnitude higher than for
SiO$_2$, and high stress Si$_3$N$_4$ could be even higher,
which is consistent with low dissipation and a long phonon mean free path.

From Table 1, we see that our fits to $C(T)$ for SiN$_{1.15\pm 0.05}$ require 
surprisingly large values of $n_o$, the TLS density of states; 
$n_0=4.5\times 10^{47} /\rm J m^3$ for 200 nm thick films and 
$n_0=1.5\times 10^{48} /\rm J m^3$ for 50
nm thick films, which are two and three orders of magnitude larger than values for
amorphous $\rm SiO_2$, respectively. This accounts for the high specific
heat below 5 K. 

Our model fits the $Q^{-1}$ data very well. At low temperatures ($T<$ 0.1 K), 
$Q^{-1}\sim T^3$, and we predict that $Q$ will increase
by up to an order of magnitude from 400 mK to 100 mK in both
stress relieved and high stress Si$_3$N$_4$. 
To obtain an upper bound for the change in barrier height due to stress, we
ignore dissipation dilution. In this case from
Table 1 we see that the mean barrier height for high stress Si$_3$N$_4$ is
$V_0=3.05\times10^4$ K $\sim 2.6$ eV which is about 33\% higher than 
$V_0=2.3\times 10^4$ K $\sim 2$ eV for stress relieved Si$_3$N$_4$. 
These values are comparable to the bond energies of Si$_3$N$_4$ \cite{Martin1987}. 
This increase in $V_0$ is consistent with our hypothesis that
stress increases the barrier heights. To see that these numbers are reasonable,
note that the difference $\Delta V_0$ in mean barrier height $V_0$ due to
stress is 7500 K. The applied stress is estimated to be about 70 K/atom. 
$n_0\times 10$ K/(N/V) in Table 1 implies that 0.06\% or 1 in 1700 atoms are
fluctuating defects. If the stress is distributed nonuniformly so that each 
atom contributes, say, 6\% of its stress to the defect, then
70 K/atom $\times$ 1700 atoms $\times$ 6\% = 7100 K $\sim\Delta V_0$.

\section{V. Dissipation of stress relieved Si$_3$N$_4$}
Why is $Q^{-1}$ in stress relieved Si$_3$N$_4$ an order of magnitude
lower than SiO$_2$? 

One might naively expect stiffer materials to have less dissipation by looking
at Eq.~(\ref{eq:Qo}) and noticing that stiffer materials will have a higher speed $v$ of
sound. This is certainly true if we compare Si$_3$N$_4$ and SiO$_2$. A measure
of the stiffness of a material is the Young's modulus $E$. 
Silicon nitride has $E=300$ GPa and $v=11.7$ km/s, while SiO$_2$ is less stiff
and has $E=66$ GPa and $v=5.8$ km/s. (We use longitudinal speeds of sound.)
However, ${\bar P} \gamma^2$ can vary from material to material and
seems to be larger in stiffer materials.
For example, PMMA is much softer than SiO$_2$ with a Young's modulus
$E$ between 1.8 and 3.1 GPa. ${\bar P}\gamma^2$ for PMMA is about an order of
magnitude smaller than the value for SiO$_2$ \cite{Yu1987}
but the values of their low temperature dissipation $Q^{-1}$ are comparable.
(${\bar P}\gamma^2 \sim 0.16\times 10^{7}$ J/m$^3$ for PMMA, and
$1.6\times 10^{7}$ J/m$^3$ for SiO$_2$.)

As another example, consider SiO$_2$ and GeO$_2$. GeO$_2$ is softer
than SiO$_2$; the Young's modulus $E$ = 45 GPa for GeO$_2$, and 
$E=66$ GPa for SiO$_2$ but
the two materials have very comparable values of the dissipation plateau at 1 K:
$Q^{-1}\sim 4\times 10^{-4}$ for GeO$_2$ and $Q^{-1}\sim 5\times 10^{-4}$ for SiO$_{2}$
\cite{Pohl2002}.
${\bar P}\gamma^{2}=1.6 \times 10^{7}$ J/m$^3$ for SiO$_2$ is double that 
of GeO$_2$ which has ${\bar P}\gamma^{2}=0.86 \times 10^{7}$ J/m$^3$ \cite{Yu1987},
while $\rho v^2 \sim 37\times 10^{6}$ J/m$^{3}$ for SiO$_2$ which is about 50\% larger
than $\rho v^2 \sim 24\times 10^{6}$ J/m$^{3}$ for GeO$_2$.
In short, the only way to determine the
correct value of ${\bar P}$ and $\gamma$ is to measure thermal conductivity,
specific heat, and dissipation for samples of silicon nitride 
with the same stoichiometry.
Stiffness alone is not enough to determine the parameters entering into the
expression for the dissipation, or to account for the reduction in dissipation
of stress relieved Si$_3$N$_4$.

So we are still left with the question of why is the dissipation of 
Si$_3$N$_4$ is an order of magnitude less than SiO$_2$.
The reason is that the atomic bonds are more
constrained in Si$_3$N$_4$. The competition between degrees of freedom and bond 
constraints is the reason why some materials are good glass-formers and 
others are not
\cite{Thorpe1983}. 
Each $m$-fold
coordinated atom provides $m/2$ constraints from fixed bond lengths, and 
$(2m-3)$ constraints from fixed bond angles \cite{Thorpe1983}. 
Since Si$_3$N$_4$ has 3 and 4-fold coordinated atoms, there are 5$\frac{4}{7}$ 
constraints per atom which exceeds the
3 degrees of freedom per atom. This is more constrained than SiO$_2$ which has 
3$\frac{2}{3}$ constraints per atom. This increase in the number of constraints
reduces the number of defects (TLS) and produces unrelieved stress that increases
the average barrier height, thus decreasing $C(T)$ and $Q^{-1}$, as well
as increasing $\kappa(T)$.

\begin{table}[!h]
\tabcolsep 0pt \caption{Parameters for SiO$_2$,
A(200nm thick SiN$_{1.15}$), B (50nm thick SiN$_{1.15}$), and Si$_3$N$_4$. 
}
\vspace*{-24pt}\label{T}
\begin{center}
\def\temptablewidth{0.5\textwidth}
{\rule{\temptablewidth}{1pt}}
\begin{tabular*}{\temptablewidth}{@{\extracolsep{\fill}}|c|l|l|l|l|}
\hline Quantities\footnotemark[1] &  $\rm SiO_2$ & $\rm A$ & $\rm B$ & 
$\rm Si_3 N_4$\footnotemark[2] \\

\hline $\rho [10^3 \rm kg/m^3]$ & $2.2$ & $2.68$ & $2.68$  & $3.18$\\
\hline $ v_{\rm L} [10^3\rm m/s] \footnotemark[3]$ & $5.8(L)3.75(T)$  
& $11.0$ & $11.7$ & $11.17$\\
\hline $\Theta_D [\rm K]$ &  $342$   & $610$ & $649$   & $446$\\
\hline $E_0  [\rm K]$   & $43$ & $76$  & $81$ & $56$\\
\hline $\bar{P} [10^{45}/\rm J m^3]$ & $0.16$ & $3$ & $10$  & $\sim 0.39$\\
\hline $S_c$ & $1300$ & $7.0$ & $2.0$ & $ 7.0*$\\

\hline $S_\kappa$ & $250$  &  $2.5$ & $1.5$  & $ 2.5*$\\
\hline $B [10^{-43}\rm s^{4}/m]$ & $1.7\times 10^4$ & $8$ &$6$  & $8*$\\
\hline $\gamma [\rm eV]$  & $2.24(L)1.73(T)$ & $5.6$ & $5.6$ &$5.6*$\\
\hline $\hbar\Omega_0 [\rm K]$ & 12 & 150 & 150 & 150/ 130 \\
\hline $2Q_0^{-1}/\pi$ $[10^{-3}]$  (L)  & $0.28$ &  $68$ & $114$  & $2.13$\\
\hline $n_0 [10^{45}/\rm J m^3]$  &  2.1 & 448 & 1490  & 58.3 /56.4 \\
\hline $n_0\times 10{\rm K}/(\mathsf{N}/\mathsf{V}) [10^{-3}] $ &$1.31\times10^{-2} $ &  1.7 & 5.6 & 
$ 0.59/0.57$\\
\hline $V_0 $ [$\times 10^4$K] & 0 & $2.3$ & $2.3$ & $2.3/3.05$\\
\hline $\sigma_0$ [$\times 10^3$ K] & $0.445$ & $9$ & $9$ & $9/7.5$ \\
 \hline
\end{tabular*}
{\rule{\temptablewidth}{1pt}}
\end{center}
\footnotetext[1]{Density $\rho$ and sound velocity $v$ are from
references \cite{Southworth2009,Queen2009}. $\Theta_D$ is calculated from $\rho$
and $v$.}

\footnotetext[2]{Parameters marked with $*$ for Si$_3$N$_4$ are estimated from 
SiN$_{1.15}$ and SiO$_2$ while ones marked with $\sim$ are estimated from other
materials.
Stress relieved and high stress values for Si$_3$N$_4$ are separated by / 
with stress relieved given first.}

\footnotetext[3]{L(T) stands for longitudinal (transverse) components. If no
data is available, we use $v_{\rm T}\approx v_{\rm L}/2$.}

\end{table}

\section{VI. Dielectric Loss}
As we mentioned in the introduction, the dielectric loss tangent $\tan\delta$
is analogous to the acoustic dissipation $Q^{-1}$. So if stress reduces
$Q^{-1}$, it should also reduce $\tan\delta$. We can estimate the effect of
stress on $\tan\delta$ using the expression in the appendix.
For SiO$_2$ with $n_ep^2=1.46\times 10^{-4}$ \cite{Schickfus1977}, $V_0=0$, and 
$\varepsilon_r=3.9$, 
$Q^{-1}_{o,dielectric}\exp\left[-V_0^{2}/(2\sigma_o^2)\right]=
Q^{-1}_{o,dielectric}\sim 5\times 10^{-4}$.
For Si$_3$N$_4$ with $\varepsilon_r=7$ \cite{Robertson2004}, and assuming 
$n_e$ is given by ${\bar P}$ in Table 1,
$p$ = 1 D, and using the values for $V_0$ and $\sigma_o$ from Table 1,
$Q^{-1}_{o,dielectric}\sim 7\times 10^{-5}$ and
$Q^{-1}_{o,dielectric}\exp\left[-V_0^{2}/(2\sigma_o^2)\right]\sim 3\times 10^{-6}$
for stress relieved Si$_3$N$_4$ and $2 \times 10^{-8}$ for high stress 
Si$_3$N$_4$. Thus stress relieved Si$_3$N$_4$ 
has the potential to lower the dielectric loss by 2 orders of magnitude, and high
stress Si$_3$N$_4$ could have dielectric loss that is up to 4 orders of magnitude lower
than SiO$_2$.

\section{VII. Alternative Model: Reduced Coupling $\gamma$ Between TLS and Phonons}
Our proposal that stress reduces the internal friction by increasing barrier heights 
can be made quantitatively consistent with the data. However, there are other possible 
explanations.  One is that stress decreases the TLS-phonon coupling $\gamma$, 
and does not change the barrier height distribution.
Figs.~\ref{fig:SpecHeat}, \ref{fig:thermCond} and \ref{fig:Q} show 
the results of this approach with $V_0=0$, $\sigma_0=9000$ K,
$\bar{P}=4.3\times10^{43}$/Jm$^3$, and $\gamma$ = 0.37 (3.96) eV for high stress
(stress relieved) Si$_3$N$_4$. The rest of the parameters are given in Table 1
for Si$_3$N$_4$. The $Q^{-1}$ fit to stress relieved Si$_3$N$_4$ is
reasonably good, but poor for high stress Si$_3$N$_4$ at high temperatures, indicating
that this model does not work as well as our hypothesis that stress increases
barrier heights if dissipation dilution plays no role.  
However, it is possible that some other set of values for the parameters could
improve the fit to the dissipation of high stress Si$_3$N$_4$.
The predicted $C(T)$ and $\kappa(T)$ resulting from 
decreasing $\gamma$ is shown in Figs.~\ref{fig:SpecHeat} and \ref{fig:thermCond}
for both high stress and stress relieved Si$_3$N$_4$.
Reducing $\gamma$ produces a thermal conductivity that is about the same
as that of the barrier height model up to about 4 K and then, at high 
temperatures, is greater than that of the barrier height model by an order of 
magnitude or more. The specific heat associated with reducing $\gamma$
is about 2 orders of magnitude lower than that of the barrier height model
at low temperatures. If stress reduces $\gamma$, the dielectric loss
will be the same for high stress and stress relieved Si$_3$N$_4$ with
$Q^{-1}_{o,dielectric}\sim 7\times 10^{-5}$.
The dielectric loss for SiO$_2$ will be the same as in the barrier height model.

The way to differentiate between these models and to determine the role of
dissipation dilution is to measure $C(T)$, $\kappa(T)$, $Q^{-1}$
and tan $\delta$ for
high stress and stress relieved Si$_3$N$_4$, and determine consistent 
values of the parameters ${\bar P}$, $\gamma$, $V_0$ and $\sigma_0$.
If dissipation dilution is the sole cause of the reduction of dissipation by
externally applied stress in high stress Si$_3$N$_4$, the thermal conductivity,
specific heat, and dielectric loss of high stress and stress relieved samples
of silicon nitride should be the same.

\section{VIII. Summary}
We have proposed three possible explanations for the reduction in dissipation
due to external and internal stress. These explanations are dissipation dilution,
stress increases the tunneling barrier $V_0$, and stress decreases 
the TLS-phonon coupling $\gamma$. 
We have used quantitative fits to show that these models are plausible.
The only way to determine the respective roles of these effects is to
determine the parameters experimentally by measuring the
dissipation, thermal conductivity, and specific heat 
on samples with the same stoichiometry. 

It is perhaps useful to view our work in the context of the history to two level systems
and glasses at low temperatures.
The original model of two level systems was proposed by Anderson, Halperin, and Varma,
and independently, by W. A. Phillips. It assumed a flat
distribution of the asymmetry energy and the tunneling barrier height of two 
level systems. This has been an enormously useful model for fitting the
low temperature thermal conductivity, specific heat, dissipation, etc. 
Tielburger, Merz, Ehrenfels, and Hunklinger used a Gaussian distribution
of the barrier height to extend the model
to fit the dissipation over a broader temperature range.
Yu and Freeman represented higher energy excitations with Einstein modes
to fit the thermal conductivity and specific heat at higher temperatures. 
Our paper moves the model forward one more step in two ways. First, we show
for the first time that one set of parameters can be used to fit 
dissipation, specific heat, and thermal conductivity at both low and
high temperatures. Second, we extend the model to include the effect of 
stress on two level systems. We propose that stress reduces the effective
number of two level systems, either by increasing the tunneling barrier height
or by decreasing the TLS-phonon coupling. As a result, stress would
decrease the dissipation
and dielectric loss as well as increases the thermal conductivity which
could have important practical applications. Examples include substrates for
integrated circuits where cooling is crucial, and superconducting qubits where
low dielectric noise is important.

\section{Acknowledgements}
We would like to thank Jeevak Parpia, Daniel McQueen, and Frances Hellman
for helpful discussions and for providing their experimental data on silicon nitride.
We thank Peter Saulson and David Cardamone for helpful discussions.
CCY thanks the Aspen Center for Physics (supported by NSF Grant
1066293) for their hospitality during which part of this paper was written. 
This work was supported in part by IARPA under grant W911NF-09-1-0368,
and by ARO grant W911NF-10-1-0494.

\section{Appendix}

In this appendix  we show the calculations involved in our estimate
of dissipation dilution. We also show the sensitivity of our fits to the 
values of the parameters.

\section{Estimation of the Contribution of Dissipation Dilution}
In dissipation dilution \cite{Huang1998} materials made stiffer by externally
applied stress without increasing their loss have a higher $Q$. 
If we write a complex Young's
modulus $E=E_0(1+i\phi)$ where $\phi$ is the internal friction and $E_0$ is a
constant, then $Q^{-1}=A\phi(\omega_0)$ where $A$ is due to dissipation
dilution and $\omega_0$ is the resonant frequency 
\cite{Saulson1990,Gonzalez1994,Huang1998}.

We can estimate the contribution of dissipation dilution to the reduction in
dissipation of high stress silicon nitride by noting that the experimental
geometry is that of a thin film square resonator \cite{Southworth2009}. 
The energy of the resonator consists of 3 parts: the kinetic energy
$K$, the energy $V_s$ from stressing the material, and the elastic
energy $V_{el}$. In this case \cite{Saulson1990}
\begin{equation}
A=\frac{V_{el}}{V_{s}+V_{el}}\approx\frac{V_{el}}{V_{s}}
\label{eq:A}
\end{equation}
since, as we shall show, $V_{el}\ll V_{s}$. We can make the approximation
\begin{equation}
\frac{V_{el}}{V_{s}}=\left(\frac{f_{0,\ \rm stress\ relieved}}{f_{0,\ \rm high\ stress}}  \right)^2
\label{eq:ratioV}
\end{equation}
where $f_0$ is the fundamental frequency of the resonator.
The energy of a square thin film resonator has 3 contributions 
\cite{Timoshenko}
\begin{eqnarray}
H&=&K+V_{s}+V_{el}\\
K&=& \frac{1}{2} \int_0^{L_x}\int_0^{L_y} \rho \left[\dot{u}(x,y)\right]^2 dx dy\\
V_s&=& \frac{1}{2} \int_0^{L_x}\int_0^{L_y} T  
\left[\left(\frac{\partial u}{\partial x}\right)^2 + \left(\frac{\partial u}{\partial y}\right)^2\right] dx dy\\
V_{el} &=& \frac{1}{2} \int_0^{L_x}\int_0^{L_y} EI  
\left[\left(\frac{\partial^2 u}{\partial x^2}
\right)^2+\left(\frac{\partial^2 u}{\partial y^2}\right)^2\right. \nonumber\\
&+&\left. 2\nu\left(\frac{\partial^2 u}{\partial x^2}\right)\left(\frac{\partial^2 u}{\partial y^2}\right)+
2(1-\nu)\left(\frac{\partial^2 u}{\partial x \partial y}\right)^2 \right] dx dy,
\end{eqnarray}
where $u(x,y)$ is the displacement perpendicular to the x-y plane, $L_x=L_y=L$ 
are the length of the sides, $\rho$ is the mass per unit area, and $E$ 
is the Young's modulus. $T$ is the tensile force per unit length given by 
$T=Sd$ where $S$ is the stress. $I=d^3/12(1-\nu)$ where $d$ is the thickness 
of the plate and $\nu$ is Possion's ratio which is $0.24$ for silicon nitride 
\cite{Gavan2009}.  The equation of motion for $u(x,y)$ is
\begin{eqnarray}
\rho \frac{d^2 u(x,y)}{d^2 t}&=&-EI \left[\frac{\partial^4}{\partial x^4}+
\frac{\partial^4}{\partial y^4}+
2 \frac{\partial^2}{\partial x^2} \frac{\partial^2}{\partial y^2} \right] u(x,y) 
\nonumber\\
&+&T\left[\frac{\partial^2}{\partial x^2}+\frac{\partial^2}{\partial y^2}\right]
u(x,y)
\end{eqnarray}
By assuming a solution of the form 
$u(x,y)=u_o\exp\left[ i k_x x+i k_y y-i\omega t\right]$, 
we find the dispersion relation: 
\begin{eqnarray}
\omega=\sqrt{\frac{T}{\rho}k^2+\frac{EI}{\rho}k^4},
\end{eqnarray}
where $k^2=k_x^2+k_y^2$. With the boundary condition 
$u(0,y)=u(L,y)=u(x,0)=u(x,L)=0$, the fundamental mode corresponds to 
$k_x=k_y=2\pi/L$ giving the fundamental resonant frequency: 
\begin{eqnarray}
f_{0}&=&\frac{\omega_{0}}{2\pi}\nonumber\\
&=&
\sqrt{\frac{2S}{\rho_0}\left(\frac{1}{L}\right)^2 +\frac{4E}{12(1-\nu^2)\rho_0}\left(\frac{2\pi d}{L}\right)^2\left(\frac{1}{L}\right)^2}
\end{eqnarray}
where $\rho_0=\rho/d$.

We can estimate the fundamental resonant frequencies of a square thin film 
resonator made of high stress and stress relieved Si$_3$N$_4$ using values from ref. 
\cite{Southworth2009}: $S$ = 1.2 GPa for high stress silicon nitride, 
$d$ = 30 nm, and $L$ = 255 $\mu$m.
The Young's modulus $E$ = 300 GPa \cite{Gavan2009}, and the mass density
$\rho_0$ = 3180 kg/m$^3$. We estimate $f_{0}\sim 3.4$ MHz for the high
stress resonator compared with the experimental value of 1.526445 MHz
\cite{Southworth2009}. For the hypothetical stress-relieved resonator with $S=0$,
we estimate $f_{0}\sim 17$ kHz. This gives a ratio of 
$\left(f_{0,{\rm high\ stress}}/f_{0,{\rm stress\ relieved}}\right)\sim 200$.
Thus from Eqs. (\ref{eq:A}) and (\ref{eq:ratioV}) 
$A\sim 2.5\times 10^{-5}$, i.e., $Q$ is enhanced up to a factor of 40,000
by dissipation dilution. However, experimentally 
\cite{Southworth2009}, the $Q$ of high stress
silicon nitride is increased by a factor of order 150 by external stress. 
The full enhancement of 40,000 is not realized probably due to external
sources of dissipation, e.g., clamping losses.
  
\section{Sensitivity of Fits to Parameters}
\begin{figure} \includegraphics[width=3.0in]{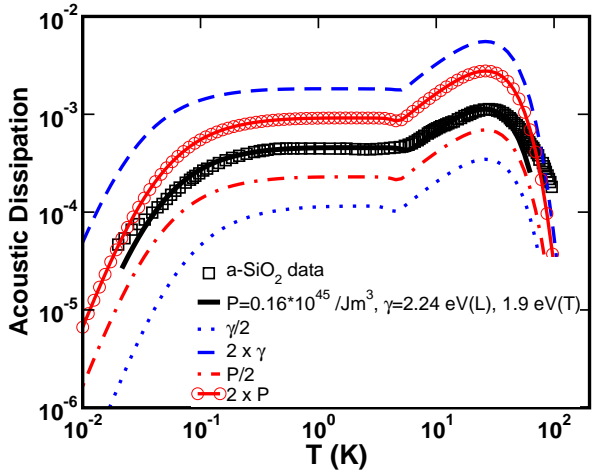}
  \caption{(Color online) 
Acoustic dissipation $Q^{-1}$ vs. temperature for various values
of the deformation potential $\gamma$ and the TLS density of states
${\bar P}$. The black squares are the SiO$_2$ data measured at 11.4 kHz from  
Tielburger {\it et al.} \cite{Tielburger1992}. The black solid line is the
fit using the values in Table 1 with ${\bar P}=0.16\times 10^{45}$ /Jm$^3$,
longitudinal $\gamma$ = 2.24 eV and transverse $\gamma$ = 1.9 eV. The 
dotted blue line comes from using values of $\gamma$ that are half as large 
while the dashed blue line comes from multiplying the values of $\gamma$ by
2. The red dot-dashed line is the result of using a value of ${\bar P}$
that is half as large, while the solid red line with open circles uses 
a value of ${\bar P}$ that is twice as large.}
\label{fig:Q_P_gamma}
\end{figure}

\begin{figure} \includegraphics[width=3.3in]{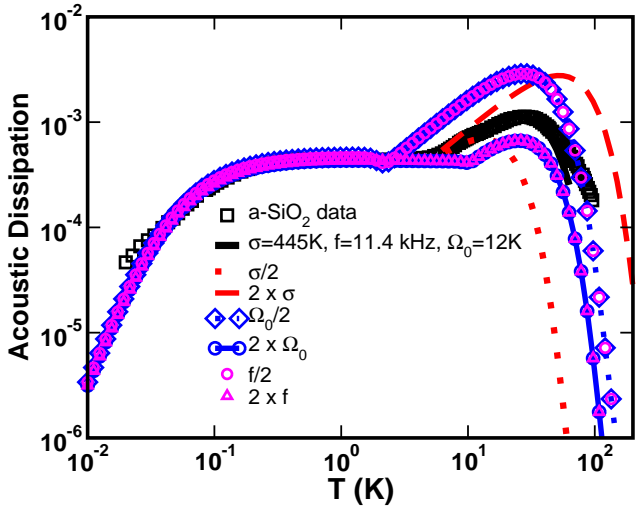}
  \caption{(Color online)
Acoustic dissipation $Q^{-1}$ vs. temperature for various values
of the width $\sigma$ of the distribution of the barrier height $V$,
the energy level spacing $\hbar \Omega_0$, and the measuring frequency
$f$. The black squares are the SiO$_2$ data measured at 11.4 kHz from
Tielburger {\it et al.} \cite{Tielburger1992}. The black solid line is the
fit using the values in Table 1 with $\sigma$ = 445 K, $\Omega_0$ = 12 K,
and $f$ = 11.4 kHz. The red dotted line uses half that value of $\sigma$, 
while the red dashed line uses twice the value of $\sigma$. The blue
dotted line with diamonds uses half the value of $\Omega_0$, and the blue 
solid line with circles uses twice the value of $\Omega_0$. The
magenta circles are for half the frequency and the magneta up triangles
are for twice the frequency $f$.}

\label{fig:Q_sigma_Omega}
\end{figure}

\begin{figure} \includegraphics[width=3.3in]{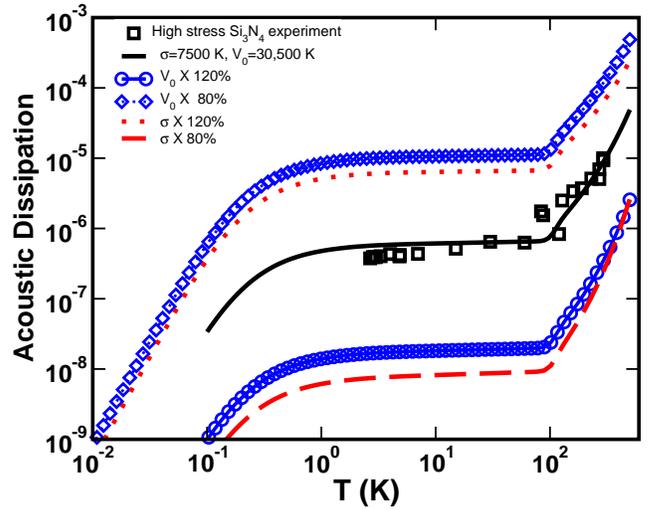}
  \caption{(Color online)
Acoustic dissipation $Q^{-1}$ vs. temperature for various values
of $\sigma$ and and $V_{0}$. The black squares are the experimental 
data points for high stress Si$_3$N$_4$ from \cite{Southworth2009}. 
The black solid
line shows the fit to the data. We have varied $\sigma$ and $V_{0}$ by 
20\% above and below the fit values to show the sensitivity of the fit 
to the values of the parameters. The values of the other parameters are
given in Table 1.}

\label{fig:Q_HighStress_Sigma_V0}
\end{figure}

To show the sensitivity of our fits to the parameters ${\bar P}$, $\gamma$, 
$\sigma$, $\Omega$, and the frequency $f$, we show in Figures
\ref{fig:Q_P_gamma}, \ref{fig:Q_sigma_Omega}, and 
\ref{fig:Q_HighStress_Sigma_V0} how the dissipation 
would change if we varied these parameters by a factor of 2 from the values
that we quoted in Table 1 in the paper.

\end{document}